# Bragg diffraction by a magnetic all-in-all-out configuration with application to a cubic cerium pyrochlore oxide

S. W. Lovesey[1, 2]

[1]ISIS Facility, STFC, Didcot, Oxfordshire OX11 0QX, UK

[2]Diamond Light Source, Harwell Science and Innovation Campus, Didcot, Oxfordshire OX11 0DE, UK

**Abstract** The Bragg diffraction of neutrons and x-rays are well-suited to the task of determining the distribution of magnetization in crystals. Applications of the two techniques proceed by contrasting observed intensities with intensities calculated with a specific model, and changing the model as need be to achieve satisfactory agreement. An all-in-all-out (AIAO) magnetic configuration of magnetic dipoles on a cubic face-centred lattice with networks of corner-sharing tetrahedra is often mentioned in the context of pyrochlore oxides, for example, but the corresponding neutron and x-ray diffraction patterns appear not have been calculated. Our results for patterns of Bragg spots from an AIAO magnetic configuration defined by a magnetic space group are symmetry informed and yield exact reflection conditions. Specifically, a long-range order of magnetic dipoles is forbidden in our model. Bulk properties arise from higher-order multipoles that include quadrupoles and octupoles. Bragg spots that exclude all magnetic multipoles other than an octupole have been discovered, and they can be observed by both neutron diffraction and resonant x-ray diffraction. All magnetic multipoles allowed in diffraction by cerium ions ($4f^1$) are presented in terms of coefficients in a well-documented and unusual magnetic ground state. Symmetry of the cerium site in the cubic structure constrains the coefficients. Our scattering amplitudes have an application in both neutron and x-ray diffraction experiments on $Ce_2Zr_2O_7$, for example, and searches for the sought-after cerium octupole. Also presented for future use is a result for the total, energy-integrated magnetic neutron scattering intensity by a powder sample.

## I. INTRODUCTION

A motivation to study Bragg diffraction by configurations of magnetic multipoles on a cubic face-centred lattice is the likely use of findings in quests to better understand pyrochlore oxides. Applications of diffraction techniques proceed by contrasting observed diffraction patterns with patterns calculated with a specific model, and refining or discarding the model according to the outcome. Symmetry informed calculations of scattering amplitudes for beams of neutrons and x-rays tuned in energy to the atomic resonance of magnetic ions are reported in our submission. In the model, magnetic ions decorate a lattice (space group $O_h^7$ or $Fd\bar{3}m$) that hosts a noncollinear all-in-all-out (AIAO) motif of magnetic dipoles depicted in Fig. 1, where dipoles point toward or away from the centre of each tetrahedron.

Magnetic cubic pyrochlore oxides of the type $A_2^{3+}B_2^{4+}O_7$ have been extensively studied in recent times. Properties of the "green fire" compounds are reviewed in several places, including Refs. [1, 2]. The structural property that A and B sites form networks of corner-sharing tetrahedra is thought to be key to many of their intriguing magnetic features. The distortion of the 8-coordinated A-site geometry from an ideal cube is very large, unlike the near perfect cubic 6-coordinated B-site geometry. Configurations of magnetic dipoles on A or B sites include the AIAO motif.

Our diffraction pattern for AIAO is applied to Kramers ions in A sites. Specifically, $Ce^{3+}$ ($4f^1$) in $Ce_2Zr_2O_7$ for which the ground state has been established [3]. Theoretical modelling of the Ce magnetic ion includes a dipole-octupole state [4], along with the possibility that the material supports a quantum spin liquid state with U(1) gauge invariance [5, 6, 7]. Our diffraction amplitudes include all Ce multipoles necessary for a meaningful characterization of long-range magnetic order, namely, a dipole (rank K = 1), an octupole (K = 3) and a triakontadipole (K = 5). The results enable us to calculate intensities of Bragg spots in terms of coefficients in the ground state. Bragg spots in the AIAO diffraction pattern that exclude all multipoles other than the cerium octupole receive particular attention with a view to future experiments.

The AIAO motif of magnetic dipoles is defined in the following section. The corresponding electronic structure factor from which we derive neutron and x-ray diffraction amplitudes appears in Eq. (7). It is the product of a magnetic multipole and a factor that couples Miller indices and angular anisotropy in the multipole. The latter are listed in an Appendix. Specific values for a Kramers Ce ion in $Ce_2Zr_2O_7$ follow from results shown in Section III [3]. Neutron and x-ray diffraction are the subjects of Sections IV-VI. A result for the total, energy-integrated neutron intensity for a powder sample appears in Section V [7]. Multipoles in neutron scattering amplitudes depend on the magnitude of the reflection vector, unlike resonant x-ray scattering amplitudes treated in Section VI. So-called neutron form factors suitable for $Ce^{3+}$ are depicted in Fig 2. The dipole form factor is strong in the forward direction [8], where octupole and triakontadipole form factors vanish [9, 10]. The stark difference in form factors can be used to advantage in distinguishing multipole contributions. Similarly, unit-cell structure factors for resonant x-ray diffraction in Section VI depend on the azimuthal angle (rotation of the crystal around the reflection vector).

## II. MAGNETIC STRUCTURE

The parent structure is taken to be cubic $Fd\bar{3}m$ (No. 227, origin choice 2). Magnetic ions occupy centrosymmetric sites 16(d) with site symmetry $D_{3d}$ ($\bar{3}m$) [1]. Principal axes ($\xi$, $\eta$, $\zeta$) are taken to be $\xi = [-1, -1, 2]/\sqrt{6}$, $\eta = [1, -1, 0]/\sqrt{2}$, and $\zeta = [1, 1, 1]/\sqrt{3}$. On rotation to cubic crystal axes (x, y, z) principal axes $\eta$ and $\zeta$ align with y and z, respectively.

An AIAO motif in Fig. 1, compatible with the irreducible representation $GM_2^+$ [11, 12], is hosted by the cubic face-centred magnetic space group $Fd\bar{3}m'$ (No. 227.131, BNS [13]). The centrosymmetric crystal class $m\bar{3}m'$ ($T_h$) allows a piezomagnetic effect, while ferromagnetism is forbidden and the only magnetoelectric effect is non-linear HEE (H and E are magnetic and electric fields, respectively). Magnetic ions occupy sites 16(d) with symmetry $\bar{3}m'$ that preserves the centre of inversion symmetry.

## III. MULTIPOLES

Axial magnetic multipoles that appear in diffraction amplitudes are denoted by $\langle t^K_Q \rangle$ or $\langle T^K_Q \rangle$ for principal axes $(\xi, \eta, \zeta)$ and cubic axes $(x, y, z)$, respectively. The multipole rank K is odd for magnetic neutron (Section IV) and resonant x-ray (Section VI) diffraction by ions in centrosymmetric sites, and the $(2K + 1)$ projections are limited to a range $-K \leq Q \leq K$ [14, 15]. In the neutron case, there is an additional caveat that the ground state is a J-manifold, otherwise even K are allowed, about which we say more in Section IV [10, 15]. Angular brackets $\langle ... \rangle$ denote the time average, or expectation value, of the enclosed spherical tensor. A triad axis of rotation symmetry $3_\zeta$ in $\bar{3}m'$ demands $Q = 3n$ in $\langle t^K_Q \rangle$, where $n$ is an integer. Values of Q are thus $Q = 0, \pm 3$ given that $K = 5$ is the maximum rank for the $f^1$ configuration [15]. For odd K, an anti-dyad $2_\eta'$ in $\bar{3}m'$ demands $(-1)^Q \langle t^K_{-Q} \rangle = \langle t^K_Q \rangle^* = \langle t^K_Q \rangle$, where the second equality follows from our definition of the complex conjugate denoted by *.

A measurement of the crystal field potential of Ce in $Ce_2Zr_2O_7$ shows that a Kramers doublet is formed from [3],

$$|u\rangle = a|J, M\rangle + b|J, -M\rangle. \tag{1}$$

Atomic states from the configuration $^2F$ in Eq. (1) have a total angular momentum $J = 5/2$ and a projection $M = 3/2$. A ground state is,

$$|g\rangle = |u\rangle + f |\bar{u}\rangle, \tag{2}$$

where $|\bar{u}\rangle$ is the time reversed version of $|u\rangle$. Normalization of $|g\rangle$ requires $\{(|a|^2 + |b|^2)(1 + |f|^2)\} = 1$. Additional restrictions on parameters a, b and f arise from site symmetry, as we shall see.

Values of $\langle t^K_Q \rangle = \langle g|t^K_Q|g\rangle$ are provided in Appendix A. The diagonal component $\langle t^K_0 \rangle = \langle t^K_\zeta \rangle$ is purely real for all a, b and f, in accord with our definition of the complex conjugate [14, 15]. And $\langle t^K_{\pm 3} \rangle$ are likewise when,

$$Z = a^*b (1 - |f|^2) + b^2 f^* - (a^*)^2 f, \tag{3}$$

is purely real.

Multipoles $\langle t^K_Q \rangle$ in neutron diffraction are strong functions of the magnitude of the reflection vector $\kappa$, and representative values of radial integrals $\langle j_c(\kappa) \rangle$ for $Ce^{3+}$ are depicted in Fig. 2. Integer $c$ is even, with $\langle j_c(0) \rangle = 0$ for $c = 2, 4, 6$, and $\langle j_0(0) \rangle = 1$. Fig. 2 features $\langle j_0(\kappa) \rangle$, $h(\kappa) = \{\langle j_2(\kappa) \rangle + (10/3) \langle j_4(\kappa) \rangle\}$ and $g(\kappa) = \{\langle j_4(\kappa) \rangle + 12 \langle j_6(\kappa) \rangle\}$ [10]. The octupole form factor $h(\kappa)$ is a maximum for $\kappa \approx 7.95$ Å$^{-1}$, where $h(\kappa)$ is seen to be much larger than $\langle j_0(\kappa) \rangle$. Most simple models of magnetic neutron scattering use $\langle j_0(\kappa) \rangle$ as an atomic form factor [7, 8].

The magnetic dipole $\langle t^1_\zeta \rangle$ aligned with the triad axis of rotation symmetry merits comments. For neutron diffraction,

$$\langle t^1_\zeta \rangle = (1/3) [\mu \langle j_0(\kappa) \rangle + (6/5) \langle L_\zeta \rangle \langle j_2(\kappa) \rangle], \tag{4}$$

where the magnetic moment $\mu = \langle L_\zeta + 2S_\zeta \rangle$ and orbital angular momentum $\langle L_\zeta \rangle = (4/3)\mu$ [15, 16, 17]. Specifically,

$$\mu = (|a|^2 - |b|^2)(1 - |f|^2) + 4(abf^*)', \qquad (5)$$

and the prime denotes the real part of the quantity $(abf^*)$. Observe that $\mu = 0$ for real a, b and f $= i$, but Z is not purely real for this choice of parameters. The necessary condition on Z is met by real a, b = a and f = 0, for example, which also results in $\mu = 0$. The null value for the magnetic moment means $\langle t^K{}_\zeta \rangle = 0$ for all K. In which case, $\langle t^K{}_{+3} \rangle$ with K = 3 and 5 alone are responsible for magnetic scattering of neutrons and x-rays. Returning to Eq. (4), a standard approximation to $\langle t^1{}_\zeta \rangle$ for neutron diffraction, referred to as the dipole approximation, is recovered from the exact result by replacing the coefficient of $\langle L_\zeta \rangle$ by unity [15, 17].

## IV. NEUTRON DIFFRACTION AMPLITUDES

An electronic structure factor,
$$\Psi^K{}_Q = [\exp(i\boldsymbol{\kappa} \cdot \mathbf{d}) \langle T^K{}_Q \rangle_\mathbf{d}], \qquad (6)$$
determines diffraction amplitudes. In Eq. (6), the reflection vector $\boldsymbol{\kappa}$ is defined by integer Miller indices $(h, k, l)$, and the implied sum is over magnetic ions in sites $\mathbf{d}$ in a unit cell. Environments in the unit cell of $Fd\bar{3}m'$ are related by pure rotations through 180° around axes x, y, and z. In consequence, $\Psi^K{}_Q(16d)$ is the same for the parent structure and the magnetic structure. Except, that is, for properties of the multipoles $\langle T^K{}_Q \rangle$, which are time-even (even K) in one case and time-odd (magnetic, odd K) in the other case. With face-centring reflection conditions even $h + k$, $h + l$, $l + k$ satisfied,

$$\Psi^K{}_Q(16d) = 4(-1)^l \langle T^K{}_Q \rangle [\{1 + \alpha\beta(-1)^Q\} - \beta\chi \exp(i\pi Q/2)\{1 + \alpha\beta^*(-1)^Q\}], \qquad (7)$$

where phase factors $\alpha = \exp(i\pi h/2)$, $\beta = \exp(i\pi k/2)$, $\chi = \exp(i\pi l/2)$. Reflection conditions depend on projections Q, and Miller indices. Notably, the conditions do not depend on K, or any specific property of an atomic multipole other than its angular anisotropy, namely, the projection Q; magnetic multipoles are relegated to fixing the size of the structure factor through parameters in the ground state and the magnitude of the reflection vector $= (2\pi/a)\sqrt{\{h^2 + k^2 + l^2\}}$ ($Ce_2Zr_2O_7$ unit cell dimension $a \approx 10.74$ Å [7]).

Bulk magnetic properties are related to the structure factor evaluated for the forward direction of scattering, i.e., $h = k = l = 0$. Under these conditions on Miller indices $\Psi^K{}_z(16d) = 0$, and $\Psi^K{}_Q(16d) = 0$ for odd Q. Thus, dipolar ferromagnetism (K = 1, Q = 0, ±1) is forbidden, in accord with the magnetic crystal class $m\bar{3}m'$. Note, however, that the bulk structure factor can be different from zero for Q = ±2, ±4. A quadrupole is striking example [10, 15, 18]. Multipoles with an even rank appear in magnetic neutron scattering amplitudes when the ground state is a mixture of J-manifolds, as predicted for $Ce_2Sn_2O_7$ [19]. A quadrupole is proportional to $\langle j_2(\kappa) \rangle$. We strive to illuminate the nature of the long-range magnetic order of octupoles and triakontadipoles through discussions of various reflection conditions derived from Eq. (7).

The electronic structure factor excludes odd Q at reflections of the type $(2n, 2n, 0)$ and $(0, 0, 2n)$. The reflection condition is odd $(m + n)$ with projections $Q = 2m$. Space-group forbidden reflections possess odd $n$, and diffraction is created by $\langle T^K_z \rangle$ and $\langle T^5_{+4} \rangle$. On the other hand, one multipole is observed with even $n$, and it is central to current discussions about the magnetic properties of cubic pyrochlore oxides [2, 4, 6]. With $\langle T^5_{+2} \rangle = 0$ in the list of multipoles in Appendix A, diffraction is created by the octupole $\langle T^3_{+2} \rangle$. Symmetry alone is responsible for this finding. The corresponding expectation value of the neutron-electron interaction operator is zero for reflections $(0, 0, 2n)$, however. The operator is usually written $\mathbf{Q}_\perp = \{\mathbf{e} \times (\mathbf{Q} \times \mathbf{e})\}$, with $\mathbf{e}$ a unit vector parallel to the reflection vector [15, 17]. Using universal expressions for the expectation value of the intermediate operator $\langle \mathbf{Q} \rangle$ reported in Ref. [15], we find $\langle \mathbf{Q}_\perp \rangle = (0, 0, \langle Q_z \rangle)$ with $\langle Q_z \rangle = -i\, 2\sqrt{210}\, \langle T^3_{+2} \rangle$ for reflections $(2n, 2n, 0)$ with even $n$. Optimum intensity occurs for $\kappa \approx 7.95$ Å$^{-1}$, which is nearly achieved for $n = 4$.

Intensity of a magnetic Bragg spot $= |\langle \mathbf{Q}_\perp \rangle|^2$ when the neutron beam is unpolarized. A polarization dependence of scattering is usually described by a departure from unity of the ratio of the reflected intensities for primary neutron beams of opposite polarization [20, 21, 22]. Such a departure in the ratio of intensities is allowed for $(2n, 2n, 0)$ with even $n$. For, nuclear and magnetic neutron amplitudes possess a like phase and interfere in diffraction, given that $\langle Q_z \rangle \propto i\, \langle T^3_{+2} \rangle$ is purely real. By contrast, there is no neutron polarization dependence in diffraction by eskolaite (chromium sesquioxide, $Cr_2O_3$), because nuclear and magnetic amplitudes are 90° out of phase. In general, a polarized neutron diffraction signal $\Delta = \{\mathbf{P} \cdot \langle \mathbf{Q}_\perp \rangle\}$, where $\mathbf{P}$ is polarization of the primary neutrons. A spin-flip intensity SF is frequently used to measure the magnetic content of a Bragg spot, with SF $= \{|\langle \mathbf{Q}_\perp \rangle|^2 - \Delta^2\}$ when $(\langle \mathbf{Q}_\perp \rangle^* \times \langle \mathbf{Q}_\perp \rangle) = 0$ [24].

The reflection condition even $n + Q$ applies to Bragg spots $(2n, 0, l)$. Setting $l = 2n'$ together with odd $n$ and odd $Q$, one finds $\langle \mathbf{Q} \rangle = (0, \langle Q_y \rangle, 0)$ and $\langle \mathbf{Q} \rangle = (\langle Q_x \rangle, 0, \langle Q_z \rangle)$ for odd and even $n'$, respectively. We consider odd $n'$ for which $\mathbf{e} \cdot \langle \mathbf{Q} \rangle = 0$ and,

$$\langle Q_y \rangle = -48\, [(1/\sqrt{2})\, \langle T^1_{+1} \rangle'' + (1/8)\sqrt{35}\, \{\sqrt{(1/15)}\, (6e_z^2 + e_x^2 - 2)\, \langle T^3_{+1} \rangle'' + e_x^2 \langle T^3_{+3} \rangle''\} \ldots ]. \quad (8)$$

Note that $\langle T^1_{+1} \rangle'' = -\sqrt{2}\, \langle T^1_y \rangle$ and the dipole form factor peaks in the forward direction. Using $\kappa \approx 1.17\sqrt{\{n^2 + n'^2\}}$ Å$^{-1}$ we observe $n = n' = 5$ yields a near optimum contribution from the octupole form factor $h(\kappa)$ in Fig. 2. For brevity of the expression, we omit triakontadipoles in the purely real amplitude $\langle Q_y \rangle$.

## V. TOTAL SCATTERING

Total intensity of neutron scattering is $\langle \mathbf{Q}_\perp \cdot \mathbf{Q}_\perp \rangle$, which we evaluate for a powder sample of a simple material. A matrix element in the expectation value of a J-manifold is,

$$\langle J, M | (\mathbf{Q}_\perp \cdot \mathbf{Q}_\perp)_{av} | J, M' \rangle = \delta(M, M')\, \{3/(2J + 1)\} \sum_K [1/(K + 1)]\, (J\|T^K\|J)^2. \quad (9)$$

A subscript "av" denotes the average over directions of the reflection vector. Reduced matrix elements $(J\|T^K\|J)$ in Eq. (9) are listed in Ref. [10]. Form factors therein are $\{\langle j_0(\kappa)\rangle + (8/5)\langle j_2(\kappa)\rangle\}$, $h(\kappa)$ and $g(\kappa)$, for K = 1, 3 and 5, respectively. Eq. (9) is a special case of the general result found in Appendix B.

## VI. RESONANT X-RAY DIFFRACTION

Magnetic multipoles can be studied by resonant x-ray diffraction, and the experimental technique has genuine advantages over neutron diffraction for the case in hand. Let us consider reflections $(2n, 2n, 0)$ and $(0, 0, 2n)$ with even $n$ that directly reveal the octupole in the AIAO magnetic configuration. For a rare earth ion, $L_2$ or $L_3$ absorption events access 4f orbitals in an electric quadrupole - electric quadrupole (E2-E2) process, with K = 0 - 4 [14, 25]. X-ray multipoles do not depend on the magnitude of the reflection vector, and both Bragg spots of interest can be studied. Magnetic multipoles in an E2-E2 absorption event have an odd rank, and the octupole denoted $\langle \tau^3_{+2}\rangle$ contributes intensity to the chosen Bragg spots. Specifically, the x-ray scattering process in which photon polarization is rotated from perpendicular to parallel to the plane of scattering, denoted $\pi'\sigma$ in a standard notation, does not contain strong charge scattering by spherical distributions of electrons [14]. An azimuthal angle $\psi$ measures rotation of the crystal sample around the reflection vector. Unit-cell structure factors for even $n$ are [26],

$$F_{\pi'\sigma}(2n, 2n, 0) = i\,\sin(\psi)\cos(\theta)\,[\cos^2(\theta) + (3\cos^2(\theta) - 2)\cos(2\psi)]\,\langle\tau^3_{+2}\rangle,$$

$$F_{\pi'\sigma}(0, 0, 2n) = \sin(2\psi)\,[\sin(\theta) + \sin(3\theta)]\,\langle\tau^3_{+2}\rangle. \tag{10}$$

The x-ray multipole $\langle\tau^3_{+2}\rangle$ is identical in form to $\langle T^3_{+2}\rangle$ in Eq. (A2). Dependence of $F_{\pi'\sigma}$ on $\psi$ and the Bragg angle $\theta$ in Eq. (10) is exact, while radial integrals and numerical factors are omitted [14, 27, 28]. For $\psi = 0$ and $(2n, 2n, 0)$ the crystal c axis is normal to the plane of scattering, and the a axis is normal to it at the origin of the azimuthal angle scan using $(0, 0, 2n)$.

The Ce $L_3$ edge energy ≈ 5.723 keV corresponds to an x-ray wave length ≈ 2.17 Å. Bragg angles in Eq. (10) at the $L_3$ edge are determined by $\sin(\theta) \approx n\,0.285$ and $\sin(\theta) \approx n\,0.202$ for $(2n, 2n, 0)$ and $(0, 0, 2n)$, respectively, meaning allowed $n = 2$ or $n = 2, 4$ depending on which reflection is observed. X-ray multipoles $\langle t^3\zeta\rangle$ and $\langle t^3_{+3}\rangle$ in $\langle\tau^3_{+2}\rangle$ are proportional to $\mu$ and Z, respectively. The coefficients depend on the Ce electronic configuration 4f[1], and the total angular momentum of the 2p core state $\gamma = 1/2$ ($L_2$) or $\gamma = 3/2$ ($L_3$). Let [14],

$$\Phi(\gamma) = +2\,(-4)\,(1/7)\,\sqrt{(1/210)}\,[-1\,(+1)\,7\,W^{(0,3)3} - 3(+3/2)\,\{-\sqrt{2}\,W^{(1,2)3} + 2\sqrt{11}W^{(1,4)3}\}]. \tag{11}$$

Here, coefficients in brackets are correct for $\gamma = 3/2$, and replace their pre-factor which is correct for $\gamma = 1/2$, e.g., the coefficient of the unit tensor $W^{(0,3)3}$ is $-2$ ($L_2$) or $-4$ ($L_3$) apart from the common factor $\sqrt{(1/210)}$. Operator equivalents for $W^{(a,b)3}$ and sum-rules for E2-E2 magnetic dichroism - linear combinations of $\Phi(1/2)$ and $\Phi(3/2)$ - have long been established

[25]. The cerium electronic configuration $4f^1$ determines the three unit tensors $W^{(a,b)K}$ in Eq. (11) and they are derived from,

$$W^{(a,b)K} = (2J + 1) \sqrt{(2K + 1)} \begin{Bmatrix} \sigma & \sigma & a \\ l & l & b \\ J & J & K \end{Bmatrix}. \qquad (12)$$

Symmetry of the 9j-symbol in Eq. (12) demands even (a + b + K) [16]. Remaining quantities are spin $\sigma = 1/2$, orbital angular momentum $l = 3$, and total angular momentum $J = 5/2$. In conclusion, structure factors in Eq. (10) for resonant x-ray diffraction are determined by cerium multipoles,

$$\langle t^3_\zeta \rangle = - (1/6) \sqrt{(7/5)} \, \mu \, \Phi(\gamma), \quad \langle t^3_{+3} \rangle = - (2/3) \sqrt{7} \, Z \, \Phi(\gamma). \qquad (13)$$

Coefficients in Eq. (13) are correct for $M = 3/2$ and $J = 5/2$ that appear in the ground state Eq. (2), and the value of $\langle \tau^3_{+2} \rangle$ in $F_{\pi'\sigma}(2n, 2n, 0)$ and $F_{\pi'\sigma}(0, 0, 2n)$ is read off from the appropriate result in the Appendix.

## VII. CONCLUSIONS

A symmetry informed analysis of Bragg diffraction by a magnetic all-in-all-out (AIAO) configuration of dipoles, depicted in Fig. 1, has produced exact reflection conditions and magnetic multipoles. Reflection conditions flow from the significant result Eq. (7) for the electronic structure factor appropriate for ions in centrosymmetric sites 16 (d) in the face-centred cubic space group $Fd\bar{3}m'$ (No. 227.131, BNS setting [13]). The AIAO configuration in question is compatible with the irreducible representation $GM_2^+$ [11, 12]. A long-range order of magnetic dipoles is forbidden. Off-diagonal components of higher-order multipoles contribute to bulk properties, including quadrupoles and octupoles visible in dichroic signals [14, 25, 29]. In diffraction experiments, reflection conditions depend on Miller indices and angular anisotropy of the magnetic multipoles. The type of conditions we find are a magnetic analogue of those derived by Templeton and Templeton in their analysis of x-ray scattering by aspherical distributions of electron charge [30, 31]. Angular anisotropy is set by site symmetry, which includes the operation of time reversal in the case of a magnetic crystal. Site symmetry and simple rotational relations between the four sites in a unit cell conspire to provide very stringent conditions on two classes of reflections that we examine in detail for the diffraction of neutrons and x-rays. All multipoles are excluded from diffraction amplitudes for these reflections apart from octupoles.

A natural setting for magnetic multipoles in a cubic crystal includes the triad axis of rotation symmetry, and compliance severely restricts the number of multipoles active in diffraction. On the other hand, multipoles in the electronic structure factor for diffraction Eq. (7) are set in cubic axes, and relations between the two types of multipoles are listed in an Appendix.

The AIAO configuration is frequently mentioned in discussions of magnetic cubic pyrochlore oxides. We applied our results to $Ce^{3+}$ and its magnetic ground state established for $Ce_2Zr_2O_7$ [3]. The two aforementioned stringent reflection conditions isolate the much-discussed cerium octupole (third-rank tensor) [4, 6]. We report an analytic expression for it in terms of coefficients in the ground state of $Ce_2Zr_2O_7$, for both neutron and resonant x-ray diffraction. The presence of nuclear scattering complicates the experimental confirmation in the case of neutron diffraction. Fortunately, a temperature dependence and critical behaviour of intensities can help separate magnetic and nuclear contributions. Likewise, polarization analysis treated in Section IV. The X-ray structure factors in Eq. (10) show a demarcation between the two classes of reflections with respect to different behaviours on rotation of the magnetic crystal about the reflection vector (an azimuthal angle scan).

**Acknowledgements** Applications of magnetic symmetry in the calculations benefited from ongoing advice from Dr D. D. Khalyavin, who prepared Fig. 1. Comments along the way from Professor A. T. Boothroyd were helpful. Professor G. van der Laan calculated radial integrals in Fig 2 using Cowan's atomic code [23].

## APPENDIX A

An expression for the magnetic dipole $\langle t^1_\zeta \rangle$ appears in Eq. (4). Higher-order multipoles allowed by $\bar{3}m'$ are,

$$\langle t^3_\zeta \rangle = -(2/5)\sqrt{(1/7)}\,\mu\,h(\kappa), \quad \langle t^3_{+3} \rangle = -(8/7)\sqrt{(1/35)}\,Z\,h(\kappa),$$

$$\langle t^5_\zeta \rangle = -(5/77)\sqrt{(5/33)}\,\mu\,g(\kappa), \quad \langle t^5_{+3} \rangle = (10/11)\sqrt{(1/231)}\,Z\,g(\kappa). \qquad (A1)$$

The the quantity Z and magnetic moment $\mu$ are defined in Eqs. (3) and (5), respectively, and Z is purely real. Multipoles with an even rank are absent because Eq. (1) contains one J-state [3]. Such multipoles represent entanglement of anapole and spatial degrees of freedom [10, 15, 18].

Turning to multipoles that appear in the electronic structure factor Eq. (7), Cartesian dipoles are obtained from $\langle T^1_x \rangle = (1/\sqrt{2})(\langle T^1_{-1} \rangle - \langle T^1_{+1} \rangle)$, $\langle T^1_y \rangle = (i/\sqrt{2})(\langle T^1_{-1} \rangle + \langle T^1_{+1} \rangle)$, $\langle T^1_z \rangle = \langle T^1_0 \rangle$ and they have the value $\langle t^1_\zeta \rangle/\sqrt{3}$, as expected. The identity $\langle T^K_{-Q} \rangle = \exp(i\pi Q/2)\langle T^K_Q \rangle$ for odd K is provided by a required invariance of multipoles with respect to the anti-dyad $2_{-xy}'$. Evidently, $\langle T^K_{+2} \rangle$ is purely imaginary $\langle T^K_{+2} \rangle = i\langle T^K_{+2} \rangle''$, or zero, while $\langle T^K_{+4} \rangle$ is purely real $\langle T^K_{+4} \rangle = \langle T^K_{+4} \rangle'$. For Q odd, $\langle T^K_{+1} \rangle = i\langle T^K_{+1} \rangle^*$ and $\langle T^K_{+3} \rangle = -i\langle T^K_{+3} \rangle^*$. Specific expressions are,

$$\langle T^3_z \rangle = -1/(3\sqrt{3})\,[2\,\langle t^3_\zeta \rangle + \sqrt{10}\,\langle t^3_{+3} \rangle], \qquad (A2)$$

$$\langle T^3_{+1} \rangle = -\exp(i\pi/4)\,(1/6)\,[\sqrt{2}\,\langle t^3_\zeta \rangle + \sqrt{5}\,\langle t^3_{+3} \rangle],$$

$$\langle T^3_{+2} \rangle = (i/3)\,[\sqrt{(5/2)}\,\langle t^3_\zeta \rangle - 2\,\langle t^3_{+3} \rangle],$$

$$\langle T^3_{+3} \rangle = \exp(-i\pi/4)\,1/(6\sqrt{3})\,[\sqrt{(10)}\,\langle t^3_\zeta \rangle + 5\,\langle t^3_{+3} \rangle],$$

$\langle T^5{}_z \rangle = -1/(6\sqrt{3})\, [\langle t^5{}_\zeta \rangle + \sqrt{70}\, \langle t^5{}_{+3} \rangle],$

$\langle T^5{}_{+1} \rangle = \exp(i\pi/4)\, 1/(6\sqrt{2})\, [\sqrt{10}\, \langle t^5{}_\zeta \rangle + \sqrt{7}\, \langle t^5{}_{+3} \rangle],$

$\langle T^5{}_{+2} \rangle = 0,$

$\langle T^5{}_{+3} \rangle = \exp(-i\pi/4)\, 1/(12\sqrt{3})\, [\sqrt{(70)}\, \langle t^5{}_\zeta \rangle - 11\, \langle t^5{}_{+3} \rangle],$

$\langle T^5{}_{+4} \rangle = -1/(12\sqrt{3})\, [\sqrt{(70)}\, \langle t^5{}_\zeta \rangle - 2\, \langle t^5{}_{+3} \rangle],$

$\langle T^5{}_{+5} \rangle = \exp(i\pi/4)\, 1/(12\sqrt{3})\, [\sqrt{14}\, \langle t^5{}_\zeta \rangle + 5\sqrt{5}\, \langle t^5{}_{+3} \rangle].$

Form factors for octupoles $\langle T^3{}_Q \rangle$ and triakontadipoles $\langle T^5{}_Q \rangle$ are $h(\kappa)$ and $g(\kappa)$, respectively, and they are depicted in Fig. 2.

### APPENDIX B

We provide a more general matrix element for total scattering $\langle J, M | \mathbf{Q}_\perp \cdot \mathbf{Q}_\perp | J', M' \rangle$ than Eq. (9), in so far that Eq. (B2) includes a unit-cell structure factor. For example,

$$\varphi_Q(K) = 4\, [(2K+1)/(K+1)]\, [3(2K-1)]^{1/2}\, (-1)^l\, [\{1 + \alpha\beta\, (-1)^Q\}$$
$$- \beta\chi \exp(i\pi Q/2)\{1 + \alpha\beta^*\, (-1)^Q\}]. \quad (B1)$$

The dependence of $\varphi_Q(K)$ on the projection Q and Miller indices $h$, $k$, $l$ is specific to the electronic structure factor Eq. (7) appropriate for $Fd\bar{3}m'$. In the present discussion, it is by way of an illustration. However, factors in Eq. (B1) involving the rank K are common to all magnetic structures.

The neutron-electron interaction operator $\mathbf{Q}_\perp$ possesses a non-trivial dependence on the reflection vector $\boldsymbol{\kappa} = \kappa\, \mathbf{e}$, with $\mathbf{e} \cdot \mathbf{e} = 1$. Its magnitude $\kappa$ appears in reduced matrix elements of magnetic multipole operators $\mathbf{T}^K$ [10, 15]. Spherical harmonics $C^K{}_Q(\mathbf{e})$ describe the orientation of $\boldsymbol{\kappa}$, chosen with normalization $\mathbf{C}^1(\mathbf{e}) = \mathbf{e}$ and a complex conjugate $C^K{}_Q(\mathbf{e})^* = (-1)^Q\, C^K{}_{-Q}(\mathbf{e})$ [32]. Using the Einstein convention for a sum on repeated indices [15],

$$\langle J, M | \mathbf{Q}_\perp \cdot \mathbf{Q}_\perp | J', M' \rangle = \langle J, M | T^K{}_\beta\, T^{K'}{}_{\beta'} | J', M' \rangle\, \varphi_\beta(K)\, \varphi_{\beta'}(K')$$
$$\times [(-1)^q\, C^{K-1}{}_\alpha\, C^{K'-1}{}_{\alpha'} \begin{pmatrix} K-1 & K & 1 \\ \alpha & \beta & q \end{pmatrix} \begin{pmatrix} K'-1 & K' & 1 \\ \alpha' & \beta' & -q \end{pmatrix} - R(K)\, R(K')\, (C^K{}_\beta\, C^{K'}{}_{\beta'})^*]. \quad (B2)$$

3-j symbols in Eq. (B2) are defined by Edmonds [32] and,

$$R(K) = (-1)^K\, \{K/[(2K-1)(2K+1)]\}^{1/2}. \quad (B3)$$

Furthermore,

$$\langle J, M|T^K_\beta T^{K'}_{\beta'}|J', M'\rangle = \langle J, M| O^x_\mu |J', M'\rangle (2x + 1)(-1)^\mu \begin{pmatrix} K' & K & x \\ \beta' & \beta & -\mu \end{pmatrix}, \quad (B4)$$

with a reduced matrix element [15, 32],

$$(J\|O^x\|J') = (-1)^{J+J'} (J\|T^K\|j)(j\|T^{K'}\|J') \begin{Bmatrix} J' & J & x \\ K & K' & j \end{Bmatrix}. \quad (B5)$$

The identity $(J\|T^K\|j) = (-1)^{j+J}(-1)^K (j\|T^K\|J)$ is valid for parity-even multipoles [33].

The powder average of total scattering is derived from,

$$\langle J, M|(\mathbf{Q}_\perp \cdot \mathbf{Q}_\perp)_{av}|J', M'\rangle = \{(K+1)/[(2K-1)(2K+1)^2]\}(-1)^\beta \varphi_\beta(K)\varphi_{-\beta}(K)$$

$$\times \langle J, M|T^K_\beta T^K_{-\beta}|J', M'\rangle. \quad (B6)$$

If $\varphi_\beta(K)$ is independent of the projection $\beta$, the sum on $\beta$ in Eq. (B6) forces $x = 0$. And,

$$(-1)^\beta \langle J, M|T^K_\beta T^K_{-\beta}|J', M'\rangle = \delta(J, J')(-1)^{j-J}(J\|T^K\|j)(j\|T^K\|J)/(2J+1). \quad (B7)$$

Eq. (9) is obtained from Eqs. (B6) and (B7) on setting $j = J$, and $\varphi_\beta(K) = \{[(2K+1)/(K+1)][3(2K-1)]^{1/2}\}$.

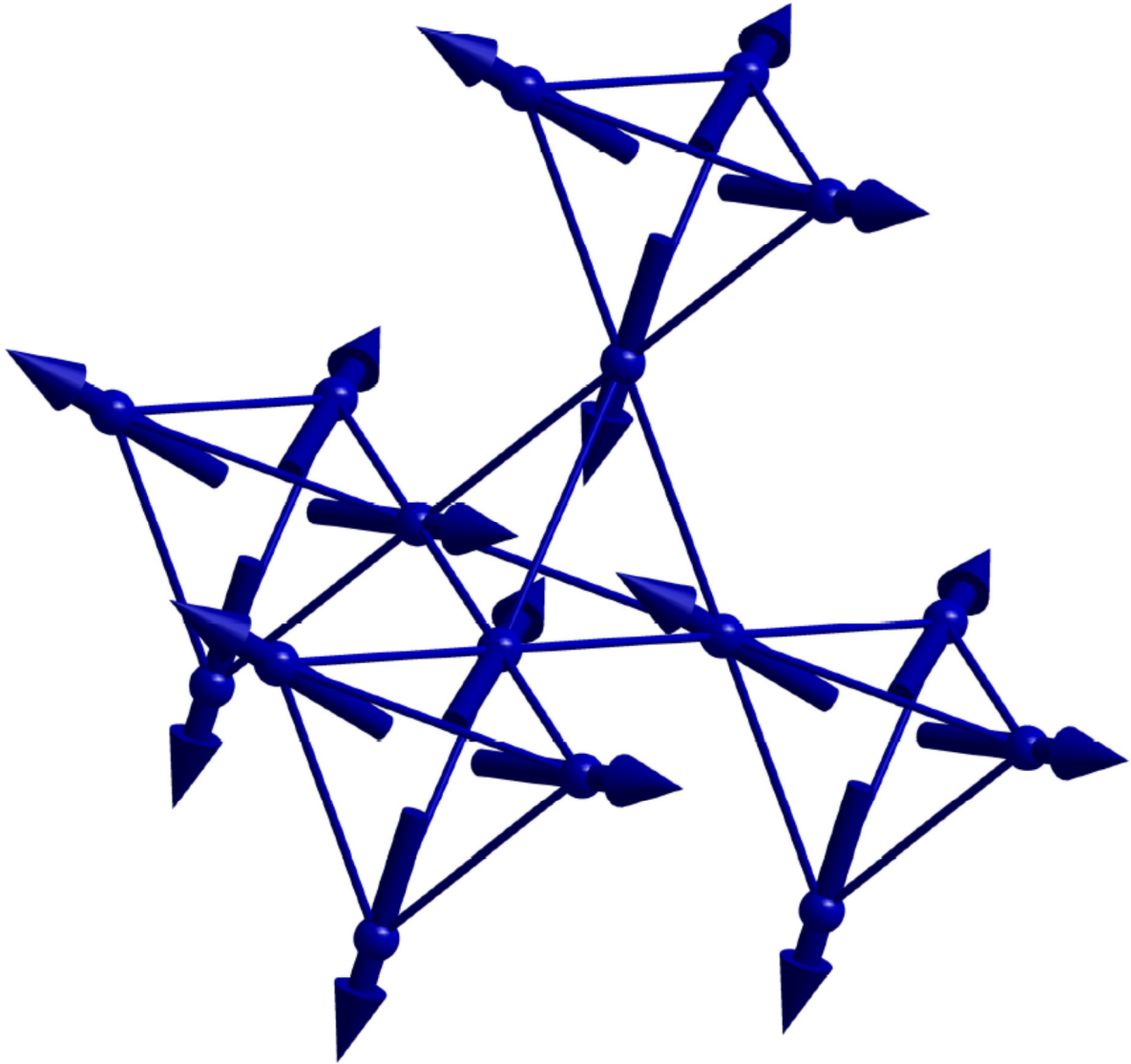

**FIG. 1**. Depiction of an AIAO configuration of magnetic dipoles compatible with sites 16 (d) in the cubic space group Fd$\bar{3}$m′ (No. 227.131, BNS [13]) and an irreducible representation GM$_2^+$ [11].

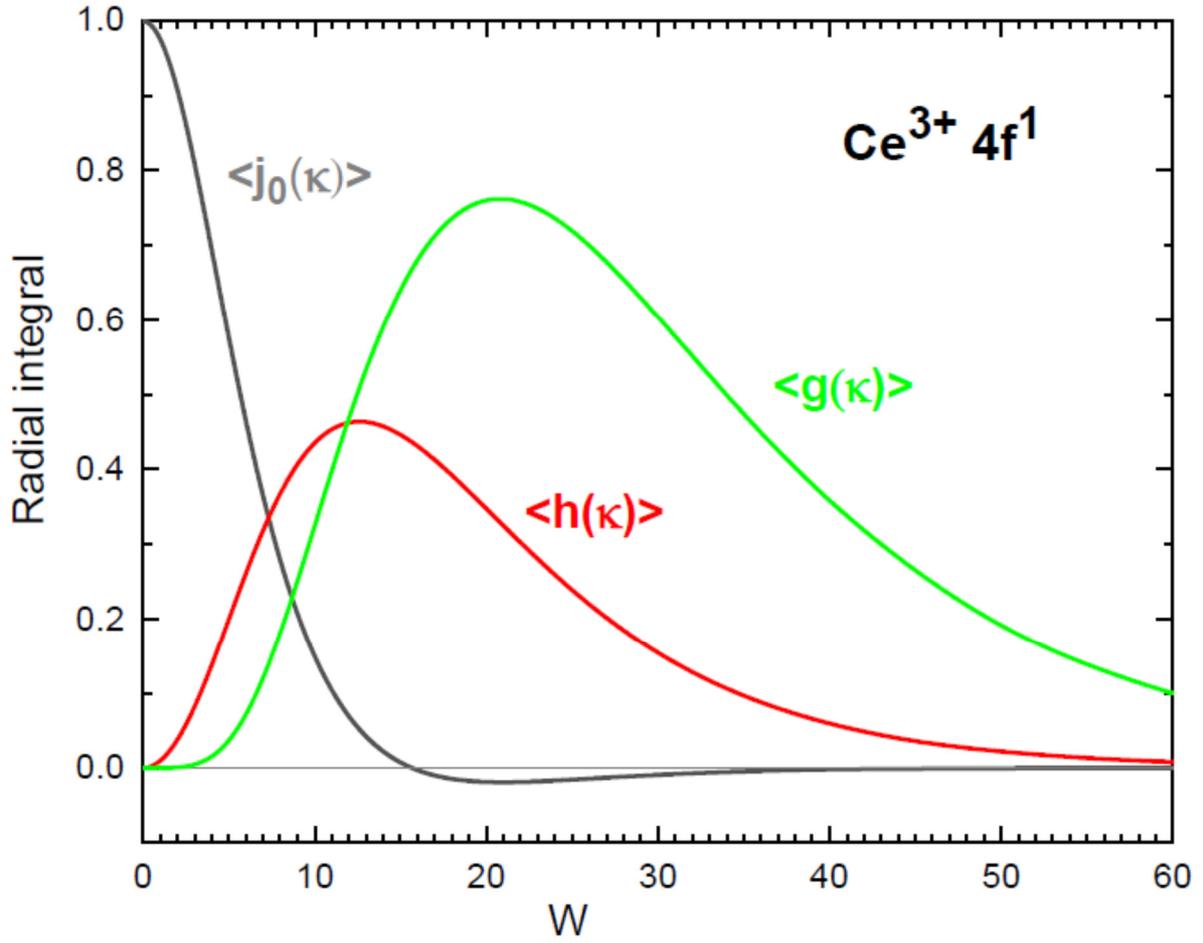

**FIG. 2**. Black, red and green curves are radial integrals $\langle j_0(\kappa)\rangle$, $h(\kappa) = \{\langle j_2(\kappa)\rangle + (10/3)\langle j_4(\kappa)\rangle\}$ and $g(\kappa) = \{\langle j_4(\kappa)\rangle + 12 \langle j_6(\kappa)\rangle\}$, respectively, for a cerium ion $4f^1$. The dimensionless parameter w and wavevector $\kappa$ are related by the Bohr radius, namely, $w = 3a_o\kappa$. Radial wave functions are obtained from Cowan's atomic code [23]. Radial integrals and they calculation are reviewed in Ref. [10].